\documentclass{kapedbk}
\usepackage{graphicx,amsmath,amsfonts,edbkps}
\normallatexbib

\begin{document}
\articletitle{
Dephasing and dynamic localization in quantum dots.}


\author{V.E.Kravtsov}
\affil{The Abdus Salam International Center for Theoretical Physics\\
strada Costiera 11, 43100 Trieste, Italy and \\
Landau Institute for Theoretical Physics, Kosygina str.,2, 117940 
Moscow, Russia.}
\email{kravtsov@ictp.trieste.it}



\begin{abstract}
The effects of dynamic localization in a solid-state system -- a 
quantum dot -- are considered. The theory of weak dynamic 
localization is developed for non-interacting electrons in a closed 
quantum dot under arbitrary
time-dependent perturbation and its equivalence to the theory of 
weak Anderson localization is demonstrated. The
dephasing due to inelastic electron scattering is shown to
destroy the dynamic localization in a closed quantum dot leading to
the classical energy absorption at times much greater than the 
inelastic scattering time. Finally a realistic case of a dot 
weakly connected to leads is studied and it is 
shown that the dynamic localization may lead to a drastic change of 
the shape of the Coulomb blockade peak in the dc conductance vs the
gate 
voltage. 

\end{abstract}



\section{Introduction}
The process of energy absorption by a quantum system with a
time-dependent Hamiltonian underlies a large part of
modern physics, both fundamental and applied.
A generic Hamiltonian can be written in the form:
\begin{equation}
\label{Ht=}
  \hat{H}(t)=\hat{H}_{0}+\hat{V}\phi(t) ,
\end{equation}
where we explicitly separated the time-independent part~$\hat{H}_{0}$
and the external perturbation~$\hat{V}$ with the time dependence specified
by a given function~$\phi(t)$.

In the textbook example of the classical Drude absorption, $\phi(t)=E(t)$
is the time-dependent electric field which is often considered to be 
a harmonic function of time $E(t)=E_{\omega} \cos(\omega t)$. Then the 
energy absorption rate $W_{0}=\frac{d{\cal E}}{dt}$ is given by the 
classical Joule heat 
formula:
\begin{equation}
\label{Joule}
W_{0}={\cal V} E_{\omega}^{2}\,\sigma(\omega),
\end{equation}
where $\sigma(\omega)$ is the frequency-dependent conductivity and 
${\cal V}$ is the volume of the system. In this classical picture the 
energy ${\cal E}$ of a {\it closed} electron system grows linearly with 
time.
Assuming the model of non-interacting electrons with the density of states
$\nu$ related by $\nu {\cal V}=1/\delta$ with the mean separation $\delta$ 
between discrete one-particle levels we find:
\begin{equation}
\label{diff}
{\cal E}={\rm const}+\int \varepsilon 
[f(\varepsilon)-f_{0}(\varepsilon)]\,\frac{d\varepsilon}{\delta}\propto
\frac{T^{2}_{\rm eff}}{\delta}+{\rm const},
\end{equation} 
where $f(\varepsilon)$ is the (non-equilibrium) electron energy 
distribution function, $f_{0}(\varepsilon)=\theta(-\varepsilon)$
is the zero-temperature Fermi-step, and $T_{\rm eff}$ is the effective
temperature of electron system. One concludes from Eq.(\ref{diff}) that
the classical absorption picture corresponds to {\it diffusion} in the 
energy space: the width of the electron energy distribution $T_{\rm eff}$
increases with time according to the diffusion law:
\begin{equation}
\label{T-diff}
T^{2}_{\rm eff}=D_{0} t,
\end{equation} 
where $D_{0}=W_{0} \delta$ is the energy diffusion coefficient. 

Apparently, this corresponds to the 
Markovian random 
walk over the spectrum of non-perturbed system caused by absorption and 
emission of energy quanta $\hbar \omega$, each absorption and emission 
events being {\it independent} of other ones. Note that in this 
random 
walk  picture the only limitation on the validity of 
Eqs.(\ref{Joule},\ref{T-diff}) is the finite width of the 
quasi-continuous energy band that leads to a saturation in 
the effective temperature or absorbed energy. The trivial example of such 
a saturation
are Rabi oscillations in a two-level system where the {\it averaged} 
energy does not change with time. 

In the past two decades attention of the scientific community was
drawn to a different and much less trivial example of {\it saturation} in 
the time-dependent energy  of {\it closed} driven systems called {\it 
dynamic localization}~(DL). In this case
the spectrum of $H_{0}$ is essentially unlimited in the energy space, 
yet,
after a certain time the absorption rate $W(t)=D(t)/\delta$ or the 
energy
diffusion coefficient $D(t)$ in Eq.(\ref{T-diff}) 
vanishes.
The DL in the energy space was
observed in numerical simulations on the kicked quantum rotor (KQR) --
particle on a circle with $\hat{H}_0=-\partial^2/\partial\theta^2$ and
$\phi(t)$~being a periodic sequence of $\delta$-pulses~\cite{Casati79},
as well as in an actual experimental realization of the KQR -- trapped
ultracold atoms in the field of a modulated laser standing 
wave~\cite{Moore}.
The mapping of the KQR to the quasi-random 1d~Anderson model has been
done in Ref.~\cite{Fishman}, a similar analogy was exploited in 
Ref.~\cite{Thouless}
to demonstrate the DL in a mesoscopic disordered
ring threaded by a magnetic flux growing linearly in time.
In Ref.~\cite{Casati90} an analogy between the KQR and band random 
matrices
was pointed out, the latter have been reduced to a 1d~nonlinear
$\sigma$~model~\cite{Mirlin}. In Ref.~\cite{Altland96} the direct
correspondence between the KQR and a 1d~nonlinear $\sigma$~model was
demonstrated.

The physical origin of {\it dynamic localization} in the energy space of 
{\it driven} systems is  essentially the same 
as for its real space counterpart -- the Anderson 
localization -- in {\it stationary} systems. This is the quantum 
nature 
of 
absorption and emission which does not reduce merely to a quantized 
step 
$\hbar\omega$ of the random walk. According to basic principles of 
quantum 
mechanics one should  consider {\it paths} in the energy space 
between an initial and a final point each of them consisting of 
many such steps. The transition probability is given by a 
square modulus of the sum of corresponding amplitudes where all 
interference terms have to be taken into account. It is these 
interference 
terms that makes the  picture of independent absorption and 
emission events incomplete and eventually may lead to dynamic 
localization. In particular the {\it weak dynamic localization} (WDL) 
exhibits 
itself as a time-dependent negative {\it correction} $\delta D(t)$ to  
the energy 
diffusion 
coefficient  which magnitude is controlled by a small 
parameter $\delta/\Gamma\ll 1$, where $\Gamma=\langle V^{2}\rangle 
/\delta$ is the typical radiation width of energy levels.

The present study consists of three parts. In the first part we develop an 
analytical theory of weak dynamic localization in a closed system of 
non-interacting electrons driven by a time-dependent 
perturbation with an {\it arbitrary} function $\phi(t)$.
To accomplish this 
goal we use a variant of the diffuson-cooperon diagrammatic 
technique
\cite{Kanzieper, Vavilov}  
in the time domain that has been developed on the basis of the Keldysh
formalism for non-equilibrium systems. We focus on the case -- most 
relevant 
for the quantum dot application -- where both the unperturbed 
system
$\hat{H}_{0}$ and the perturbation operator $\hat{V}$ are described 
by the 
Gaussian ensemble of random matrices or by an equivalent random 
field Hamiltonian considered in the zero-mode 
approximation. Such description has proved 
[see \cite{Kanzieper} and references therein] to be 
valid for low-energy domain of diffusive quantum dots and is believed to 
be also accurate for ballistic quantum dots with irregular 
boundary.
We establish an intimate relation between WDL and
dephasing by time-dependent perturbation using the notion of {\it 
no-dephasing points} first introduced in Ref.\cite{Wang}. 
We show that the existence and character of weak dynamic localization 
depends crucially on the spectral and symmetry properties of 
$\phi(t)$.
Furthermore we show that by merely changing $\phi(t)$ one can 
obtain the 
dependence of $\delta D(t)$ that reproduces the dependence of 
WL correction to electric  conductivity $\delta 
\sigma(t_{\varphi})$ on the dephasing time $t_{\varphi}$ in quasi-one 
dimensional wires,  two- and three-dimensional systems.

In the second part we give a brief description of the effect of 
{\it electron-electron interaction} on dynamic localization in {\it 
closed} systems assuming
that in the absence of interaction the {\it strong} dynamic localization
occurs at $t>t_{*}$ with the effective
temperature at saturation $T_{*}=\sqrt{D_{0}t_{*}}$. We show that in 
contrast to the 
Anderson localization 
where the hopping conductance is finite but {\it exponentially small} at 
low temperatures, the strong dynamic localization is destroyed by 
electron-electron interaction even in the case where the characteristic 
time of inelastic electron-electron scattering $t_{\rm 
ee}=t_{\rm 
ee}(T_{*})$
is much larger than the localization time $t_{*}$. What is left of DL
in {\it closed} systems with electron interaction is a suppression of 
the 
energy absorption rate $W(t)/W_{0}\sim t_{*}/t_{\rm ee}$ at 
intermediate times, 
$t_{*}< t 
< t_{\rm ee}$.  
For times longer than $t_{\rm ee}\ln(t_{\rm ee}/t_{*})$ the 
classical absorption rate $W_{0}$ is again recovered.

Finally, in the third part we address a realistic case of a quantum dot in 
a {\it steady-state} regime
under time-dependent perturbation where both electron-electron interaction
and {\it electron escape} into leads are taken into account. We show that 
a 
signature of dynamic localization can be observed in an almost closed 
quantum dot with the escape rate $\gamma_{\rm esc}\ll \delta$. This 
is a 
plateau at the tail of the Coulomb blockade peak in the dc conductance 
vs. the gate voltage. We obtain an analytic expression for the shape of 
the 
Coulomb blockade peak in driven quantum dots and identify a region 
of parameters where the plateau can occur.
The main results of the first two parts are published in 
Ref.\cite{BSK,Bas} and a detailed study of the Coulomb blockade 
regime in a driven quantum dot is presented in \cite{KrBas}.
\section{Weak dynamic localization}
The goal of this section is to show how the cooperon-diffuson diagrammatic 
technique \cite{GLKh} which is the main theoretical tool for describing 
weak Anderson 
localization (WAL) and mesoscopic phenomena, can be extended as to 
include 
non-equilibrium processes 
considered in the time-domain. This consideration is based on the Keldysh 
technique \cite{Keld} and is described in detail 
Ref.\cite{Kanzieper,Vavilov}. Here we
present only the main practical hints which allow to derive the WDL 
correction along the same lines as those used to derive WAL 
corrections to 
conductivity \cite{GLKh}.

We start by the standard \cite{Keld} expression for the energy 
distribution function
\begin{equation}
\label{f-h}
f(\varepsilon,t)\equiv\frac{1}{2}-\frac{1}{2}\int 
h\left(t+\frac{\tau}{2},t-\frac{\tau}{2}\right)\,e^{-i\tau\varepsilon}\,d\tau, 
\end{equation}
in terms of the Keldysh Green's function
${\cal G}^{K}(t,t')$:
\begin{equation}
\label{h-K}
{\cal G}^{K}(t,t')=-2\pi i\nu\,  h(t,t').
\end{equation}
Then the time-dependent absorption rate  
$W(t)=D(t)/\delta$  which is related to the 
time-dependent energy diffusion 
coefficient $D(t)$, is given by: 
\begin{equation}
\label{rate}
D(t)=-\frac{1}{2\nu}\,\frac{\partial }{\partial t}\lim_{\tau\rightarrow 
0}\frac{\partial}{\partial\tau}\,{\cal 
G}^{K}\left(t+\frac{\tau}{2},t-\frac{\tau}{2}
\right).
\end{equation}
Next we note that the Keldysh Green's function for the free electron gas
coupled with an external time-dependent field by 
${\cal H}_{\rm ef}(t)=\hat{V}\,\phi(t)$ can be expressed in terms of retarded 
${\cal G}^{R}$ and 
advanced ${\cal G}^{A}$ Green's functions as follows:
\begin{eqnarray}
{\cal G}^{K}(t,t')&=&\int 
dt_{1}dt_{1}'\;{\cal 
G}^{R}(t,t_{1})\,{\cal
G}^{A}(t_{1}',t')\;h_{0}(t_{1}-t_{1}')\,[{\cal 
H}_{\rm ef}(t_{1}')-{\cal H}_{\rm ef}(t_{1})]\nonumber \\ \label{K-RA} 
&+& \int dt_{1} h_{0}(t-t_{1})\,{\cal G}^{R}(t_{1},t')-\int 
dt_{1} 
{\cal 
G}^{A}(t,t_{1})\,h_{0}(t_{1}-t'),
\end{eqnarray}
where $h_{0}(t)$ is the Fourier-transform 
of the equilibrium distribution 
function $\tanh(\varepsilon/2T)=1-2f_{0}(\varepsilon)$ which is supposed 
to hold in the absence of perturbation ${\cal H}_{\rm ef}$. 
Note also that
the functions ${\cal G}^{R(A)}(t,t')$ take account of the
external time-dependent field {\it in all orders} in ${\cal H}_{\rm ef}$
and therefore depend on two time
arguments and not only on their difference.

The 
perturbation operator
in Eq.(\ref{K-RA}) is not necessarily random and without loss of 
generality it can be assumed to be traceless $Tr{\cal H}_{\rm ef}=0$. 
For instance one can consider the perturbation by a time-dependent 
space-homogeneous electric field ${\bf E}(t)=-\partial_{t} {\bf 
A}={\bf E}_{0}\phi(t)$, where 
${\cal H}_{\rm 
ef}= -e\hat{{\bf v}} 
{\bf A}(t)$ with zero average of the electron velocity ${\bf 
v}$ over the Fermi surface. 
\subsection{Diffuson-Cooperon diagrammatic technique in the time domain}

Now we assume that the Hamiltonian $\hat{H}_{0}$ of the unperturbed 
system is that of free electrons in a Gaussian random impurity field
$U({\bf r})$ which results in the diffusive electron dynamics with 
a small
elastic mean free path $\ell\ll L$ compared to the system size $L$.
Averaging over disorder realizations $U({\bf r})$ can be done using 
impurity diagrammatic technique \cite{AGD}. For relatively weak external 
fields 
such that 
\begin{equation}
\label{cond1}
eE_{0}\ell\ll \hbar\omega
\end{equation}
the essentially nonlinear effect 
of the field is on 
the two-particle correlation functions --diffusons and 
Cooperons-- while 
for the 
{\it averaged} single-particle Green's function an expansion in 
${\cal H}_{\rm ef}$ up to the second order is sufficient:
\begin{eqnarray}
\langle  {\cal 
G}^{R(A)}(t,t')\rangle &=&\delta(t-t')\,\left[G^{R(A)}+G^{R(A)}\,{\cal 
H}_{\rm 
ef}(t)\,G^{R(A)}\right.\nonumber\\ 
\label{G} 
&+& \left. G^{R(A)}\,{\cal 
H}_{\rm
ef}(t)\,G^{R(A)}\,{\cal H}_{\rm ef}(t)\,G^{R(A)} \right]. 
\end{eqnarray}
In Eq.(\ref{G}) it is assumed that the elastic scattering time 
$\tau_{0}=\ell/v_{F}$ is the smallest relevant time scale, and 
\begin{equation}
\label{G-xi}
G^{R(A)}(\xi_{\bf p})=\frac{1}{\xi_{{\bf
p}}\mp i/2\tau_{0}}, 
\end{equation}
where $\xi_{{\bf 
p}}=\frac{p^{2}}{2m}-\varepsilon_{F}$. 

Using Eq.(\ref{G}) one can decouple the time- and momentum- 
integrations. 
The latter for  a closed loop of $G^{R(A)}$ functions can be  
done as 
a simple 
pole 
integral $\nu \int d\xi\, G^{R(A)}(\xi)...G^{R(A)}(\xi)$, as in the 
stationary case. 
In particular,
one can repeat the standard derivation  of the diffusons and 
Cooperons
as the corresponding ladder series of the impurity diagrammatic 
technique.

The diffuson
\begin{eqnarray}
\label{D}
&&\langle {\cal G}^{R}({\bf r},{\bf
r'};t_{+},t^{\prime}_{+})
{\cal G}^{A}({\bf r'},{\bf r};t_{-}^{\prime},t_{-})
\rangle\\ \nonumber
&=&2\pi\nu\,\delta(\tau-\tau')\,
e^{i{\bf rK}_{d}(t,\tau)}D_{\tau}(t,t';{\bf r},{\bf r'})e^{-i{\bf 
r'K}_{d}(t',\tau)},
\end{eqnarray}
where $t_{\pm}=t\pm\tau/2$, $t'_{\pm}=t'\pm\tau'/2$, ${\bf 
K}_{d(c)}(t,\tau)={\bf 
A}(t+\tau/2)\mp {\bf A}(t-\tau/2)$,
obeys the equation \cite{Kanzieper}:
\begin{eqnarray}
\left\{\frac \partial {\partial
t}-D\nabla_{{\bf r}}^{2} - i{\bf r}{\bf E}_{0}\left[
\phi\left(t+
\frac \tau 2\right)-\phi\left(t-\frac \tau 2\right)\right]
\right\} \\ \nonumber 
\times
D_{\tau}(t,t^{\prime };{\bf r},{\bf r'})= \delta (t-t^{\prime
})\delta({\bf r}-{\bf r'})
\label{diF}
\end{eqnarray}
where $D=v_{F}\ell/d$ is the diffusion coefficient in the $d$-dimensional 
space.

The corresponding equation for the Cooperon 
\begin{eqnarray}
\label{C}
&&\langle {\cal G}^{R}({\bf r},{\bf
r'};t_{+},t^{\prime}_{+})
{\cal G}^{A}({\bf r},{\bf r'};t_{-},t_{-}^{\prime})\rangle\\ \nonumber
&=&2\pi\nu\,\delta(t-t')\,
e^{i{\bf r K}_{c}(t,\tau)} C_{t}(\tau,\tau';{\bf r},{\bf r'})e^{i{\bf r' 
K}_{c}(t,\tau')}
\end{eqnarray}
reads \cite{Kanzieper}:
\begin{eqnarray}
\left\{2\frac \partial {\partial
\tau}-D\nabla_{{\bf r}}^{2} - i{\bf r}{\bf E}_{0}\left[
\phi\left(t+
\frac \tau 2\right)-\phi\left(t-\frac \tau 2\right)\right]
\right\} \cr
\times
C_{t}(\tau,\tau^{\prime };{\bf r},{\bf r'})= 2\delta (\tau-\tau^{\prime
})\delta({\bf r}-{\bf r'})
\label{cooP}
\end{eqnarray}
Equations (\ref{diF},\ref{cooP}) should be supplemented with the Neumann 
boundary conditions:
\begin{equation}
\label{bound}
\frac{\partial D_{\tau}(t,t';{\bf r},{\bf r}')}{\partial n}=0,\,\,\,\,
\frac{\partial C_{t}(\tau,\tau';{\bf r},{\bf r}')}{\partial n}=0.
\end{equation}
In contrast to the initial  transverse gauge, in the longitudinal 
gauge we switched to in Eqs.(\ref{D}--\ref{cooP}) the 
boundary 
conditions do not depend on the time-dependent perturbation.

Without time-dependent 
perturbation Eqs.(\ref{diF},\ref{cooP}) are just simple diffusion 
equations
which have a complete set of stationary solutions $\varphi_{\mu}({\bf r})$ 
such that  $D\nabla_{{\bf r}}^{2}\varphi_{\mu}({\bf 
r})=E_{\mu}\varphi_{\mu}({\bf r})$ including 
the zero mode $\varphi_{0}={\rm const}$. One can use this set as a 
convenient basis for solving Eqs.(\ref{diF},\ref{cooP}) in the presence 
of
time-dependent perturbation \cite{Kanzieper}. In the limit
\begin{equation}
\label{cond}
eE_{0}L, \,\hbar\omega \ll E_{Th},
\end{equation}
where $E_{Th}=\hbar E_{1}\sim \hbar D/L^{2}$ is the Thouless energy, the 
zero mode makes the main contribution, other modes should be treated as 
perturbations. Then the second-order perturbation theory yields for the 
{\it zero-mode} diffuson and Cooperon 
\cite{Kanzieper,Vavilov,AAKh}:
\begin{equation}
\label{0-diff}
D_{\tau}(t,t^{\prime })=\theta(t-t')\,\exp\left\{-\Gamma 
\int^{t}_{t'}
\left[\phi\left(t''+\frac{\tau}{2}\right)-\phi\left(t''-\frac{\tau}{2}\right)
\right]^{2}\,dt''\right\},
\end{equation}
\begin{equation}
\label{0-coop}
C_{t}(\tau,\tau^{\prime })=\theta(\tau-\tau')
\,\exp\left\{-\frac{\Gamma}{2} \int^{\tau}_{\tau'}
\left[\phi\left(t+\frac{\eta}{2}\right)-\phi\left(t-\frac{\eta}{2}\right)
\right]^{2}\,d\eta\right\},
\end{equation}
where 
\begin{equation}
\label{ga}
\Gamma=\sum_{\mu\neq 0} \frac{[e{\bf r}_{0\mu}{\bf 
E_{0}}]^{2}}{E_{\mu}}\sim \frac{(eE_{0}L)^{2}}{E_{Th}}
\end{equation}
and ${\bf r}_{0\mu}=\int {\bf dr}\, \phi_{0}\,{\bf r}\phi_{\mu}({\bf r})$.

The validity condition for the zero-mode approximation can be obtained 
using Eqs.(\ref{cond1},\ref{cond},\ref{ga}):
\begin{equation}
\label{cond-f}
\frac{\ell}{L}\,\sqrt{\frac{\Gamma}{E_{Th}}}\ll\frac{\hbar\omega}{E_{Th}}\ll 
1.
\end{equation}

It is remarkable that the same expressions  
for the diffuson and the 
Cooperon can be obtained  \cite{Kanzieper,Vavilov} if one starts from the 
Hamiltonian Eq.(\ref{Ht=}) with $\hat{H_{0}}$ and $\hat{V}$ taken from 
the Gaussian Orthogonal ensemble of real symmetric $N\times N$ random 
matrices with 
the probability distributions:
\begin{equation}
\label{proba}
{\cal P}_{H_{0}}\propto \exp\left[ 
-\frac{\pi^{2}\,tr\,H_{0}^{2}}{4N\delta^{2}}\right],\;\;\;\;{\cal P}_{V}
\propto \exp\left[
-\frac{\pi\,tr\,V^{2}}{4\Gamma\delta}\right].
\end{equation}
This demonstrates that for small conductors (quantum dots) and relatively 
weak external fields where Eq.(\ref{cond-f}) satisfies, the 
{\it non-random} 
time-dependent perturbation 
$V_{0}\phi(t)=e\hat{{\bf r}}{\bf E}_{0}\,\phi(t)$ in a {\it random} system 
is equivalent to a 
{\it random-matrix} perturbation $\hat{V}$ with $\langle 
\hat{V}^{2}\rangle\sim V_{0}^{2}/g$, where $g=E_{Th}/\delta \gg 1$. This 
statement is believed to hold not only for diffusive quantum dots but also 
for generic ballistic quantum dots.

However, perhaps the simplest way to generate a correct diffuson-cooperon 
diagrammatic technique in the zero-dimensional limit Eq.(\ref{cond-f}) is 
to 
consider in Eq.(\ref{Ht=}) 
\begin{equation}
\label{fict}
\hat{H}=\frac{\hat{p}^{2}}{2m}+U({\bf r})+V({\bf r})\phi(t)
\end{equation}
where $U({\bf r})$ and $V({\bf r})$ are {\it independent} Gaussian random 
fields with zero mean values and the following correlation functions:
\begin{equation}
\label{corr-f}
\langle U({\bf r})\,U({\bf 
r}')\rangle=\frac{1}{2\pi\nu\tau_{0}}\,\delta({\bf r}-{\bf r}'),\,\;
\langle V({\bf r})V({\bf r}') \rangle=\frac{\Gamma}{\pi\nu}\delta({\bf 
r}-{\bf r}');\;\;\;\;\Gamma \tau_{0}\ll 1. 
\end{equation}
We will use this representation below because it corresponds to a minimal
deformation of the impure-metal Hamiltonian that is a starting point of
a conventional impurity diagrammatic technique and because it helps to 
avoid 
subtle effects of the boundary conditions Eq.(\ref{bound}) which are 
important 
for a 
deterministic global perturbation.

\subsection{One loop DL correction}
As a first step of averaging over disorder in Eq.(\ref{K-RA}) we  
single out the ladder series that represent diffusons and 
Cooperons. As usual \cite{GLKh} the diagrams are classified 
with respect to the number of diffuson or Cooperon loops. The zero- and 
one-loop diagrams are shown in Fig.1a and Fig.1c, respectively.
The next step is to properly average the remaining products of retarded 
and advanced single-particle Green's functions (denoted by $R$ and $A$ in 
Fig.1). So we obtain the Hikami boxes denoted by the shadowed polygons
in Fig.1b and Fig.1d. 
\begin{figure}[ht]
\vskip.2in
\includegraphics[width=.7\textwidth]{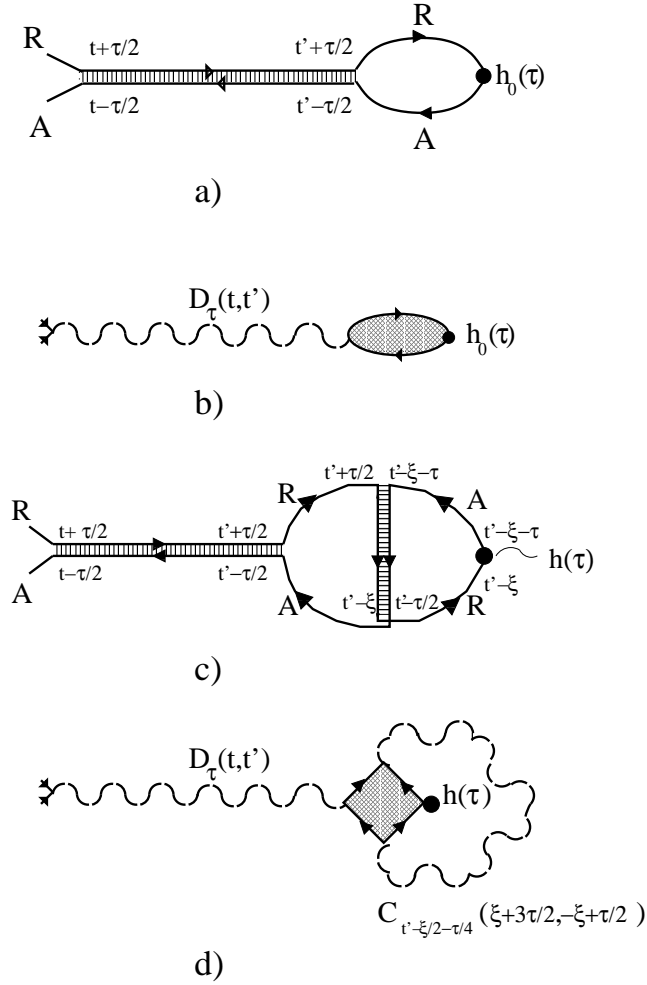}
\caption{Diagrams for the Keldysh Green's function ${\cal 
G}^{K}(t+\tau/2,t-\tau/2)$: 
a). the zero-loop diagram describing the classical absorption; the 
diffuson is 
given by a ladder series, the retarded or advanced Green's 
functions are depicted by 
solid lines b). the same 
diagram with Hikami boxes shown by shadowed loop and a diffuson denoted 
by a wavy line; c). one loop WDL correction with a diffuson and a 
Cooperon given by ladder series; d). one loop WDL correction with 
the Hikami box denoted by a shadowed rhomb; the diffuson and the loop 
Cooperon are denoted by the wavy lines} 
\end{figure}
\noindent
In order to compute the Hikami boxes, one substitutes Eq.(\ref{G}) for 
the 
averaged retarded or advanced Green's functions in Fig.1a,c retaining 
only terms second order  in the perturbation operator ${\cal H}_{\rm 
ef}=V({\bf r})\,\phi(t)$. Then one performs averaging over the random 
field $V({\bf r})$. It is shown in Fig.2 as the dashed-dotted line 
connecting two vertices. In addition to that rules of the impurity 
diagrammatic 
technique require additional averaging over the impurity field $U({\bf 
r})$ which is shown by the dotted line in Fig.2.  
\begin{figure}[ht]
\vskip.2in
\includegraphics[width=1.0\textwidth]{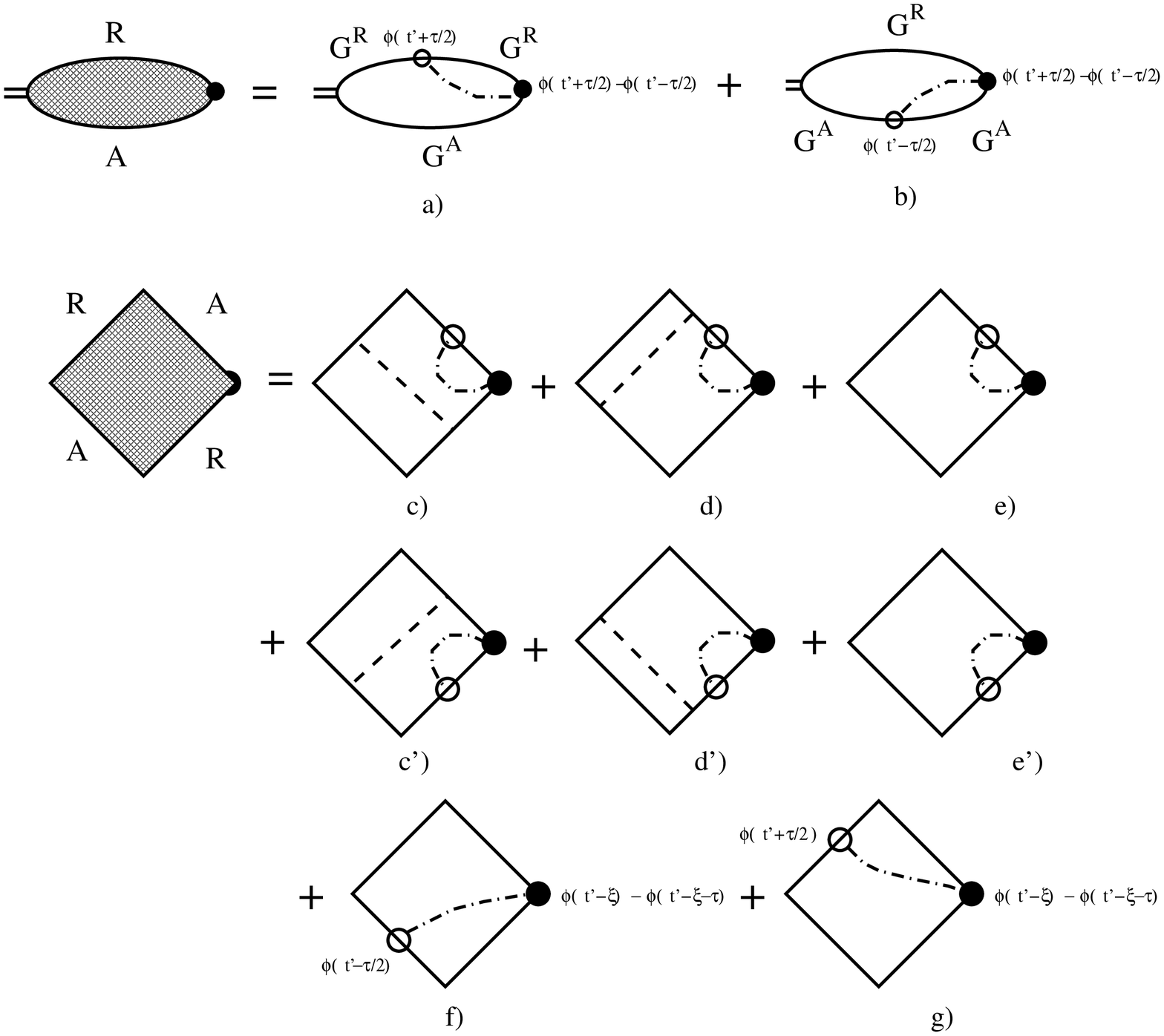}
\caption{Hikami boxes: solid lines denote retarded or 
advanced Green's functions, Eq.(\ref{G-xi}); the dashed-dotted line
and the dotted line represent the $\langle V \,V \rangle$ 
and the $\langle U\,U \rangle$ correlator Eq.(\ref{corr-f}), 
respectively. The black circle denotes the $V\phi(t)$ vertex 
in Eq.(\ref{K-RA}); the open circle corresponds to the $V\phi(t)$ 
vertex in Eq.(\ref{G}).}
\end{figure}
\noindent
The further calculation of the Hikami boxes is standard and reduces to 
pole integrals of $G^{R(A)}(\xi)$ functions Eq.(\ref{G-xi}). 

One can 
easily 
see that the diagrams Fig.2a and 
Fig.2b
have opposite signs.
As the result the Hikami box Fig.2a,b in the zero-loop diagram Fig.1b  
is proportional to 
$[\phi(t'+\tau/2)-\phi(t'-\tau/2)]^{2}$,
so that the function $h(t+\tau/2,t-\tau/2)$ in Eq.(\ref{h-K}) obeys
in the zero-loop approximation the following equation:
\begin{equation}
\label{diff-Eq}
\left\{\frac{\partial}{\partial t} +
\Gamma\,\left[\phi\left(t+\frac{\tau}{2}\right)-\phi\left(t-\frac{\tau}
{2}\right)\right]^{2}\right\}\,h\left(t+\frac{\tau}{2},t-\frac{\tau}{2}\right)
=0.
\end{equation}
If one is interested in  
the energy resolution $\Delta\varepsilon\gg \omega$, where $\omega$ is a 
typical period of a time-dependent field, it suffices to
consider $\omega \tau \ll 1$. Then from Eq.(\ref{diff-Eq},\ref{f-h}) we 
obtain 
diffusion in the energy space: 	
\begin{equation}
\label{diff1-Eq}							
\left\{\frac{\partial}{\partial t} -
\Gamma\,(\partial_{t}\,\phi(t))^{2}\,
\frac{\partial^{2}}{\partial\varepsilon^{2}}
\right\}\,f(\varepsilon,t)
=0.
\end{equation}
If in addition we are interested in the time resolution $\Delta t\gg 
\omega^{-1}$, the time-independent diffusion coefficient can be defined:
\begin{equation}
\label{D-coeff}
D_{0}=\Gamma\,\overline{(\partial_{t}\,\phi(t))^{2}}\sim \Gamma\omega^{2},
\end{equation}
where $(\partial_{t}\phi)^{2}$ 
averaged over 
time 
interval much larger than the typical period of oscillation is supposed 
to be well defined and independent of time.

One can see from Eq.(\ref{diff1-Eq}) that the parameter $\Gamma$ defined 
in Eqs.(\ref{ga},\ref{proba},\ref{corr-f}) has a physical meaning of
an inverse time of making a step $\sim \omega$ in a random walk 
over the 
energy space.

The correction $\delta D(t)$ to the energy diffusion coefficient is 
described by the diagrams Fig.1c,d where the Hikami box is given by 
diagrams Fig.2c-g. One can easily see that in 
the zero-mode approximation (zero external momentum in any vertex 
of the rhomb)
three diagrams Fig.2c-e cancel each other. The same is valid for 
the diagrams Fig.2c'-e'. The remaining diagrams Fig.2f,g have opposite 
signs, so that the entire Hikami box in Fig.1d is proportional to
$(\phi(t'+\tau/2)-\phi(t'-\tau/2))\,(\phi(t'-\xi)-\phi(t'-\xi-\tau))$.
Then Eq.(\ref{rate}) yields \cite{BSK}: 
\begin{equation}
\label{WDLC}
\delta 
D(t)=\frac{\Gamma\delta}{\pi}\int_{0}^{t}\partial_{t}\phi(t)\,
\partial_{t}\phi(t-\xi)\,C_{t-\xi/2}(\xi,-\xi)\,d\xi,
\end{equation}
where $C_{t-\xi/2}(\xi,-\xi)$ is the zero-mode Cooperon given by 
Eq.(\ref{0-coop})
and $t$ is time passed since the onset of the time-dependent perturbation
$\phi(t)$.
Note that the correction $\delta D(t)$ is a purely quantum 
mesoscopic effect that vanishes if the mean level spacing $\delta$
between energy levels of the non-perturbed system is equal to zero.

Equation (\ref{WDLC}) that describes the {\it weak dynamic localization} 
is the main result of this section.  Below we establish some of its 
implications.
\subsection{No-dephasing points and dynamic localization}
Let us consider the simplest example of a harmonic perturbation 
$\phi(t)=\cos \omega t$. Then for $\omega \xi\gg 1$ one can write:
\begin{equation}
\label{coop-gamma}
C_{t-\xi/2}(\xi,-\xi)=\exp\left\{-2\xi\, \Gamma\, 
\sin^{2}\left(\omega t-\frac{\omega\xi}{2} 
\right)\right\}.
\end{equation}
A remarkable phenomenon is that the negative exponential that describes 
{\it 
dephasing} by a harmonic perturbation vanishes on a certain set of 
{\it no-dephasing} points:
\begin{equation}
\label{NDP}
\omega\xi_{k}=2\omega t -2\pi k,\;\;\;\; k=0,\pm1,\pm2,...
\end{equation}
At large $\Gamma\xi$ only a close vicinity of these points contributes to 
the 
integral Eq.(\ref{WDLC}). Expanding $\sin^{2}$ around no-dephasing points
and performing the Gaussian integration we arrive at:
\begin{equation}
\label{sum}
\frac{\delta D(t)}{D_{0}}=-\frac{\delta}{\omega}\,
\sum_{\xi_{k}} \sqrt{\frac{2}{\pi\Gamma\,\xi_{k}}}\approx -\sqrt{\frac{t}{t_{*}}},
\end{equation}
where
\begin{equation}
\label{loctime}
t_{*}=\frac{\pi^{3}\Gamma}{2\delta^{2}}.
\end{equation}
This result is valid at $t\ll t_{*}$ and $\xi\sim t \gg 1/\Gamma$ which is only 
compatible if
\begin{equation}
\label{ga-de}
\Gamma\gg \delta.
\end{equation}
Remarkably, the WDL correction Eq.(\ref{sum}) increases with time.
Equation (\ref{sum}) sets a time scale $t_{*}$ where the negative 
WDL correction 
is of the 
order of the classical energy diffusion coefficient and one may expect 
the strong 
dynamic localization.

Note that it is only because of the no-dephasing points which are 
zeros of the 
dephasing function 
\begin{equation}
\label{deph-func}
\Gamma_{c}(t)=\frac{\Gamma}{2}\,\overline{\left[\phi\left(t+\frac
{\eta}{2}\right)-
\phi\left(t-\frac{\eta}{2}\right)
\right]^{2}},
\end{equation}
that the correction $\delta D(t)/D_{0}$ grows 
with time and may become of order one despite a small parameter 
$\delta/\Gamma$.
For instance, the white-noise perturbation $\phi(t)$ such that 
the time average $\overline{\phi(t)\phi(t')}\propto \delta(t-t')$, 
results in a 
constant 
dephasing function $\Gamma_{c}=\Gamma\,\overline{\phi^{2}}$.  
Then the WDL correction Eq.(\ref{WDLC}) is  finite $\delta D(t)\sim 
D_{0}\,\delta/\Gamma$ at $t\rightarrow \infty$
and small compared to $D_{0}$.

In order to understand the physical meaning of no-dephasing condition 
we consider 
two electron trajectories in {\it real space} with loops traversed in 
opposite 
directions (see Fig.3), the traversing time being $t$. Interference 
between such 
trajectories is known to be a 
cause of 
the weak Anderson localization. Let us assume that a time-dependent 
vector-potential $A(t)$ is present. Then the phase difference between 
trajectories 
1 and 2 in Fig.3 becomes random:
\begin{equation}
\label{phase}
\delta\varphi=\int_{0}^{t} A(t')\;[v_{1}(t')-v_{2}(t')]\,dt'.
\end{equation}
Using the condition $v_{1}(t')=-v_{2}(t-t')$ we obtain
\begin{equation}
\label{phase1}
\delta\varphi=\int_{-t/2}^{t/2}[A(t/2+t')+A(t/2-t')]\,v_{1}(t'+t/2)\,dt'.
\end{equation}
If the typical period of oscillations in $A(t)$ is large compared to the 
velocity correlation time, one obtains after averaging over directions 
of 
velocity 
$v_{1}$:
\begin{equation}
\label{phase2}
\langle(\delta\varphi)^{2}\rangle\propto 
\int_{-t/2}^{t/2}[A(t/2+t')+A(t/2-t')]^{2}\,dt'.
\end{equation}
For a harmonic $A(t)=\sin\omega t$ it is easily seen that the phase 
difference is 
zero for trajectories of any shape provided that the traversing 
time $t$ is 
synchronized with the period of the field:
\begin{equation}
\label{syn} 
t=\frac{2\pi k}{\omega},\;\;\;k=1,2,..
\end{equation}
that is, the dephasing is absent
for trajectories which traversing time 
is equal to a {\it multiple of the period} 
of perturbation.
\begin{figure}[ht]
\vskip.2in
\includegraphics[width=0.6\textwidth]{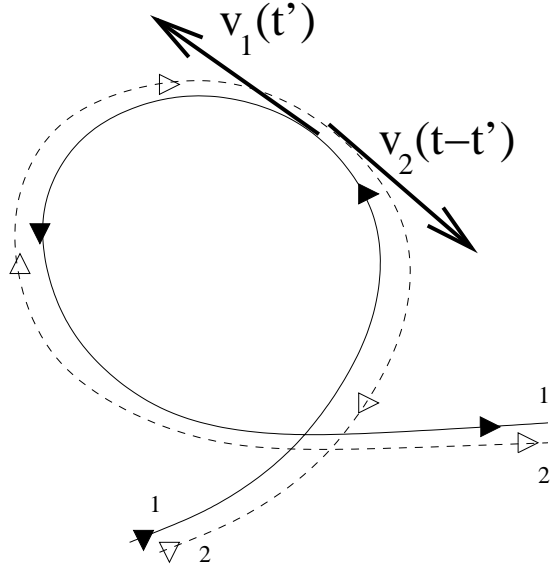}
\caption{Time-reversal conjugate trajectories.}
\end{figure}
\noindent
\subsection{No-dephasing points and generalized time-reversal 
symmetry}
Now let us consider a {\it generic periodic} perturbation:
\begin{equation}
\label{period}
\phi(t)=\sum_{m=1}^{\infty}A_{n}\,\cos(m \omega\, t+\varphi_{m}).
\end{equation}
First of all we require the diffusion coefficient 
Eq.(\ref{D-coeff}) to be finite:
\begin{equation}
\label{finite}
D_{0}=\frac{\Gamma}{2}\,\sum_{m=1}^{\infty}m^{2}\,A_{m}^{2}<\infty.
\end{equation}
This is only so when $A^{2}_{m}$ decreases faster than $1/m^{3}$. 
In particular, it is {\it not} the case when $\phi(t)$ is the 
periodic 
$\delta$-function as in the KQR model 
\cite{Casati79, Fishman}. In similar cases the sum 
Eq.(\ref{finite}) should be cut off at $m\sim E_{Th}/\omega$, so 
that the energy diffusion coefficient $D_{0}$ grows with increasing 
$E_{Th}$ or the size of matrix $N\sim E_{Th}/\delta$ in the 
equivalent random matrix model Eq.(\ref{proba}). 

The cooperon dephasing function Eq.(\ref{deph-func}) that 
corresponds to 
Eq.(\ref{period}) 
takes the 
form:
\begin{equation}
\label{deph-m}
\Gamma_{c}(t)=\Gamma\sum_{m=1}^{\infty}A_{m}^{2}\,\sin^{2}(m\omega 
t+\varphi_{m}).
\end{equation}
\begin{figure}[ht]
\vskip 1.0in
\includegraphics[width=0.8\textwidth]{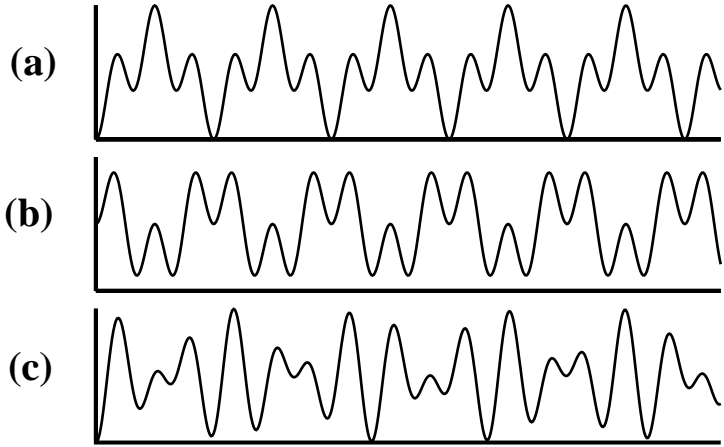}
\caption{
The Cooperon dephasing function $\Gamma_{c}(t)$ for
(a)~periodic $\phi(t)$ obeying Eq.~(\ref{TR}): a regular array
of
zeros; (b)~generic periodic $\phi(t)$: a gap;
(c)~quasi-periodic $\phi(t)$ with two incommensurate frequencies:
a pseudo-gap.}
\end{figure}
\noindent

The no-dephasing points $t_{k}$ determined by $\Gamma_{c}(t_{k})=0$ 
exist only provided that
phases of all harmonics are sinchronized: $\varphi_{m}=m\,\varphi$ 
(see Fig.4a).
This condition implies that the perturbation $\phi(t)$ can be 
made a symmetric function by a shift of the origin 
$\tilde{\phi}(t)\equiv \phi( 
t-\varphi/\omega)$:
\begin{equation}
\label{TR}
\tilde{\phi}(t)=
\tilde{\phi}(-t).
\end{equation}
Apparently, this is a generalization of the {\it time-reversal}
condition for the case of time-dependent perturbation and for
quantities that survive averaging over time intervals much larger 
than the period of oscillations in $\phi(t)$. It has been first 
established in Ref.\cite{Wang}.

If this condition is violated (see Fig.4b) and $\min 
\Gamma_{c}(t)=\gamma>0$, 
the Cooperon $C_{t}(\xi,-\xi)$
acquires an exponentially decreasing factor $e^{-2\gamma\,\xi}$
that limits the growth of WDL correction for $t\gamma\gg 1$.
For a generic case $\gamma\sim\Gamma$ we have $\delta 
D(t)/D_{0}\sim 
(\delta/\Gamma)\ll 1$ and the {\it one-loop} WDL correction given 
by Eq.(\ref{WDLC}) can be neglected.

The analogous situation arises in the weak Anderson 
localization when the time-reversal symmetry is violated by
a constant magnetic field or magnetic impurities. In this case the 
one-loop WAL correction also gradually vanishes leading to the 
anomalous magneto-resistance. The difference  
is that in the case of WDL the Hamiltonian remains {\it real}, 
represented e.g. by the 
{\it orthogonal ensemble} of random matrices. Yet the time-reversal 
symmetry can be  broken just by time-dependence of perturbation.
From this perspective the harmonic perturbation is an important 
{\it exception} rather than a paradigm of a periodic perturbation.   

The analogy with the WAL suggests that the main WDL correction
in the case where the condition Eq.(\ref{TR}) is violated is given 
by the two-loop diagrams containing only diffusons. Indeed, let us 
consider the diffuson dephasing function:
\begin{equation}
\label{deph-diff} 
\Gamma_{d}(\tau)=\Gamma\,\overline{\left
[\phi\left(t''+\frac{\tau}{2}\right) 
-\phi\left(t''-\frac{\tau}{2}\right) 
\right]^{2}}.
\end{equation}
For $\phi(t)$ given by Eq.(\ref{period}) it reduces to:
\begin{equation}
\label{ddf}
\Gamma_{d}(t)=2\Gamma \sum_{m=1}^{\infty} A_{m}^{2}\,\sin^{2}\left( 
m\,\frac{\omega\tau}{2}\right).
\end{equation}
As was expected, $\Gamma_{d}(t)$ is independent of phases 
$\varphi_{m}$ and has a set of no-dephasing points $\tau_{k}=2\pi 
k/\omega$ {\it for any} periodic perturbation Eq.(\ref{period}). As 
the result, the two-loop WDL correction is a growing 
function of time \cite{SkvDenKr}:
\begin{equation}
\label{TLC}
\frac{\delta^{(2)}D(t)}{D_{0}}=-\frac{\pi}{24}\,\frac{t}{t_{*}},
\end{equation}
where
\begin{equation}
\label{loct}
t_{*}=\frac{\pi^{3}\,\Gamma\,\overline{m^{2}}}{2\delta^{2}}
\end{equation}
and 
$\overline{m^{2}}=\sum_{m=1}^{N}m^{2}\,A_{m}^{2}/
\sum_{m=1}^{N}A_{m}^{2}$.
\subsection{Incommensurability of frequencies, avoided no-dephasing 
points and weak Anderson localization in higher dimensions.}

Now let us consider the perturbation $\phi(t)$ as a sum of several 
incommensurate harmonics:
\begin{equation}
\label{inc}
\phi(t)=\sum_{m=1}^{d}A_{m}\,\cos(\omega_{m}t+\varphi_{m}).
\end{equation}
In this case the Cooperon dephasing function $\Gamma_{c}(t)$ 
is given by:
\begin{equation}
\label{incCDF}
\Gamma_{c}(t)=\Gamma\sum_{m=1}^{d}A_{m}^{2}\,\sin^{2}
(\omega_{m}t+\varphi_{m}),
\end{equation}
where all $\omega_{m}$ do not have a common multiplier.
In this case one has a {\it distribution} of {\it avoided}
no-dephasing points as in Fig.4c.

In order to compute the one-loop WDL correction Eq.(\ref{WDLC})
we expand {\it each} of the exponentials [sf. 
Eq.(\ref{coop-gamma})] 
$\exp\left \{ 
-2\xi\,\Gamma\,A_{m}^{2}\,\sin^{2}\left(\omega_{m}t-
\frac{\omega_{m}\xi}{2}+\varphi_{m} 
\right)\right \}$ in a Fourier series using:
\begin{equation}
\label{Fourier}
e^{z\,\cos(2\omega_{m}t+2\varphi_{m})}=\sum_{n_{m}=-\infty}^{+\infty}
I_{n_{m}}(z)
\,e^{in_{m}(2\omega_{m}t+2\varphi_{m})},
\end{equation}
where $I_{n_{m}}(z)$ is the Bessel function.

Then Eq.(\ref{WDLC}) reduces to:
\begin{eqnarray}
\label{SUM}
\delta
D(t)&=&-\frac{\Gamma\delta}{2\pi}\int_{0}^{t}d\xi\,
\sum_{n_{1}=-\infty}^{+\infty}...\sum_{n_{d}=-\infty}^{+\infty}
\sum_{m=1}^{d}
\omega_{m}^{2}A^{2}_{m}\,(\ln
I_{n_{m}}(z_{m}))^{'}\,
\nonumber
\\&\times&
\prod_{m=1}^{d}
I_{n_{m}}(z_{m})\,e^{-z_{m}}\,
e^{i n_{m}(2\omega_{m}t-\omega_{m}\xi+2\varphi_{m})},
\end{eqnarray}
where $z_{m}=\xi\Gamma A_{m}^{2}$ and the prime denotes the 
derivative $d/dz_{m}$.

In order to obtain the behavior of $\delta D(t)$ averaged over time 
intervals much larger than the typical period of oscillations one 
should take into account only $n_{m}$ obeying the constraint:
\begin{equation}
\label{constr}
\sum_{m=1}^{d} n_{m}\,\omega_{m}=0.
\end{equation}
It is exactly the condition of {\it incommensurability}
of frequencies $\omega_{m}$ that the constraint Eq.(\ref{constr}) 
can be satisfied 
only if {\it all} $n_{m}=0$. One can see that at such a condition
$\delta D(t)$ does not depend on the phases $\varphi_{m}$ and is 
given by:
\begin{equation}
\label{SUM1}
\delta
D(t)=-\frac{\Gamma\delta}{2\pi}\int_{0}^{t}d\xi\,\sum_{m=1}^{d}
\omega_{m}^{2}A^{2}_{m}\,(\ln 
I_{0}(z_{m}))^{'}\,\prod_{m=1}^{d}
I_{0}(z_{m})\,e^{-z_{m}}
\end{equation}
Further simplification is possible if we consider $A_{m}=1$ and 
$\Gamma t\gg 1$. In this limit $(\ln
I_{0}(z_{m}))^{'}=1$, $I_{0}(z_{m})\approx 
e^{z_{m}}/\sqrt{2\pi z_{m}}$ and one arrives at \cite{BSK}:
\begin{equation}
\label{d-dim}
\frac{\delta D(t)}{D_{0}}=-\frac{\delta}{\pi \Gamma}\int_{\sim 
1}^{\Gamma t}\frac{dz}{(2\pi z)^{d/2}}.
\end{equation}  
Remarkably, the dependence on $t$ in Eq.(\ref{d-dim}) 
coincides with the dependence of WAL corrections to conductivity of 
a $d$-dimensional conductor on 
the dephasing time $\tau_{\varphi}$. In the case of two 
incommensurate harmonics the WDL correction grows 
with time
logarithmically:
\begin{equation} 
\label{2d}
\frac{\delta D(t)}{D_{0}}=-\frac{\delta}{2\pi^{2} 
\Gamma}\,\ln(t\Gamma).
\end{equation}
For three and more incommensurate harmonics $\delta D(t)/D_{0}\sim 
\delta/\Gamma$ is small. It is yet to be studied whether or not
a dynamic localization {\it transition} is possible for $d\geq 3$ 
at some 
$\Gamma_{\rm crit}\sim \delta$ as it is claimed to be the case
for the KQR \cite{Shepel}.
\section{Role of electron interaction on dynamic localization in 
closed quantum dots.}
So far we considered energy absorption in a closed system of {\it 
non-interacting 
fermions} under time-dependent perturbation.  Although the 
many-body character of the system was 
important for making the diffusion in the energy space a mechanism 
of energy {\it absorption} due to the presence of the filled Fermi 
see,
inelastic processes of interaction were completely disregarded.
We have demonstrated how the weak dynamic localization arises in 
such 
systems and how much does it resemble the Anderson localization 
(AL) 
corrections to conductivity. This analogy makes us believe that
WDL corrections for periodic perturbations evolve with time to 
produce the strong DL in the 
same way as WAL corrections lead to a vanishing conductance of 
low-dimensional systems of sufficiently large size in the absence 
of dephasing mechanisms. So we {\it assume} that in a quantum dot 
without 
electron interaction at times $t\gg 
t_{*}$ the energy 
diffusion 
coefficient becomes {\it exponentially small}. By this time the 
electron temperature rises from a low fridge 
temperature to the effective temperature $T_{*}\sim \omega 
\Gamma/\delta$.

The question is what happens if electron interaction is present.
More precisely, what happens  if the dephasing rate 
$\gamma_{\varphi}=1/t_{\rm ee}$ 
related with 
interaction is small but finite:
\begin{equation}
\label{deph-rate}
0<\gamma_{\varphi}(T_{*})\ll \frac{1}{t_{*}}.
\end{equation}
In the problem of strong AL the conductance $\sigma$ remains 
exponentially small provided that the phase-breaking length
is much greater than the localization length $L_{\varphi}\gg 
L_{\rm loc}$ and the temperature is much smaller than the mean 
level spacing in the localization volume $\delta_{\rm loc}=(\nu 
L_{\rm loc}^{d})^{-1}$. The first condition ensures localization of 
eigenstates. The second one makes hopping over 
pre-formed localized states an exponentially  rare event, so that 
$\sigma\sim e^{-(\delta_{\rm loc}/T)^{\alpha}}$ with some positive 
exponent $\alpha$.

The point of a crucial difference between the strong AL and 
the strong DL is that in the latter it is the inverse localization 
time $t_{*}^{-1}\sim \delta^{2}/\Gamma$ 
that 
plays a role of $\delta_{\rm loc}$. In the case 
when $\delta$ is the smallest energy scale we have
\begin{equation}
\label{T-t}
T_{*}\sim \frac{\omega\Gamma}{\delta}\gg \frac{1}{t_{*}}.
\end{equation} 
As a result, no 
exponentially small factor arises in the hopping over 
the Floquet 
states 
in contrast to the hopping over localized states in  a real space.

The qualitative picture that follows from the consideration 
\cite{Bas} based 
on the Fermi Golden Rule is the following. A given electron
absorbs energy from the external field during the time $t_{*}$
until the strong DL develops itself. At $t>t_{*}$ the dynamically 
localized state with the energy $T_{\rm eff}\sim 
T_{*}$ above the Fermi sea is formed which is characterized by the 
exponentially small absorption rate. So it continues until after a 
time $t_{\rm ee}$ a 
collision with another electron breaks the phase-tuning necessary 
for DL. The  "efficiency" of each collision in the destruction of 
DL is close to 100\% because the typical energy transfer 
is large $T_{*}\gg 
1/t_{*}$.  Then the absorption rate jumps to a classical 
value and it takes another $t_{*}$ to reach DL for the second time,
but now with energy $2T_{*}$. So, for a given electron relatively 
short time intervals $\sim t_{*}$
of absorption are followed by larger periods $\sim t_{\rm ee}$ 
of 
DL,
the energy increasing by $T_{*}$ after each circle of absorption.  
The averaged absorption rate $W_{\rm in}$ is given by:
\begin{equation}
\label{AvAR}
W_{\rm in}=W_{0}\,\frac{t_{*}}{t_{*}+t_{\rm ee}}.
\end{equation}
Were electron-electron collision time $t_{\rm ee}$ energy 
independent, the averaged absorption rate would 
stay constant
at all times $t>t_{*}$. However,
$1/t_{\rm ee}$ typically increases with increasing the energy 
$T_{\rm eff}$, so 
does 
also
the averaged absorption rate. Then we obtain from Eq(\ref{diff}):
\begin{equation}
\label{Rate}
\frac{d T^{2}_{\rm eff}}{dt}=D_{0}\,\frac{t_{*}}{t_{*}+t_{\rm 
ee}(T_{\rm 
eff})}.
\end{equation}  
The time $t_{\rm ee}$ that corresponds to inelastic collisions 
of 
electrons in a quantum dot with 
the energy transfer $\sim T_{\rm eff}\ll E_{Th}$ has been 
calculated 
in Ref.\cite{SIA}:
\begin{equation}
\label{SIA}
\frac{1}{t_{\rm ee}(T_{\rm eff})}=\delta\,\left(\frac{T_{\rm 
eff}}{E_{Th}}\right)^{2}.
\end{equation}
For $t_{*}\ll t_{\rm ee}(T_{\rm eff})$ Eqs.(\ref{Rate}, 
\ref{SIA}) 
suggest 
that the effective 
electron temperature increases exponentially:
\begin{equation}
\label{effT}
T_{\rm eff}=T_{*}\,\exp\left[\frac{t}{2t_{\rm 
ee}(T_{*})} \right]
\end{equation}
rather than remaining a  constant $\sim T_{*}$ as it does for 
non-interacting electrons.
The corresponding absorption rate
$W_{\rm in}(t)$ takes the form:
\begin{equation}
\label{W-in}
W_{\rm in}(t)=W_{0}\,\left(\frac{t_{*}}{t_{\rm 
ee}(T^{*})}\right)\,\exp\left[\frac{t}{t_{\rm
ee}(T_{*})} \right]
,\,\,\,\,\,\,t>t_{*}.
\end{equation}
The exponential growth of effective temperature continues until
$t_{\rm ee}(T_{\rm eff})$ decreases down to $t_{*}$. At larger 
times $t>t_{\rm ee}(T_{*})\,\ln(t_{\rm ee}(T_{*})/t_{*})$ the 
classical absorption rate is restored: $W_{\rm in}=W_{0}$.
The sketch of the time-dependence of absorption rate is shown in 
Fig.5.
\begin{figure}[ht]
\vskip.2in
\includegraphics[width=0.8\textwidth]{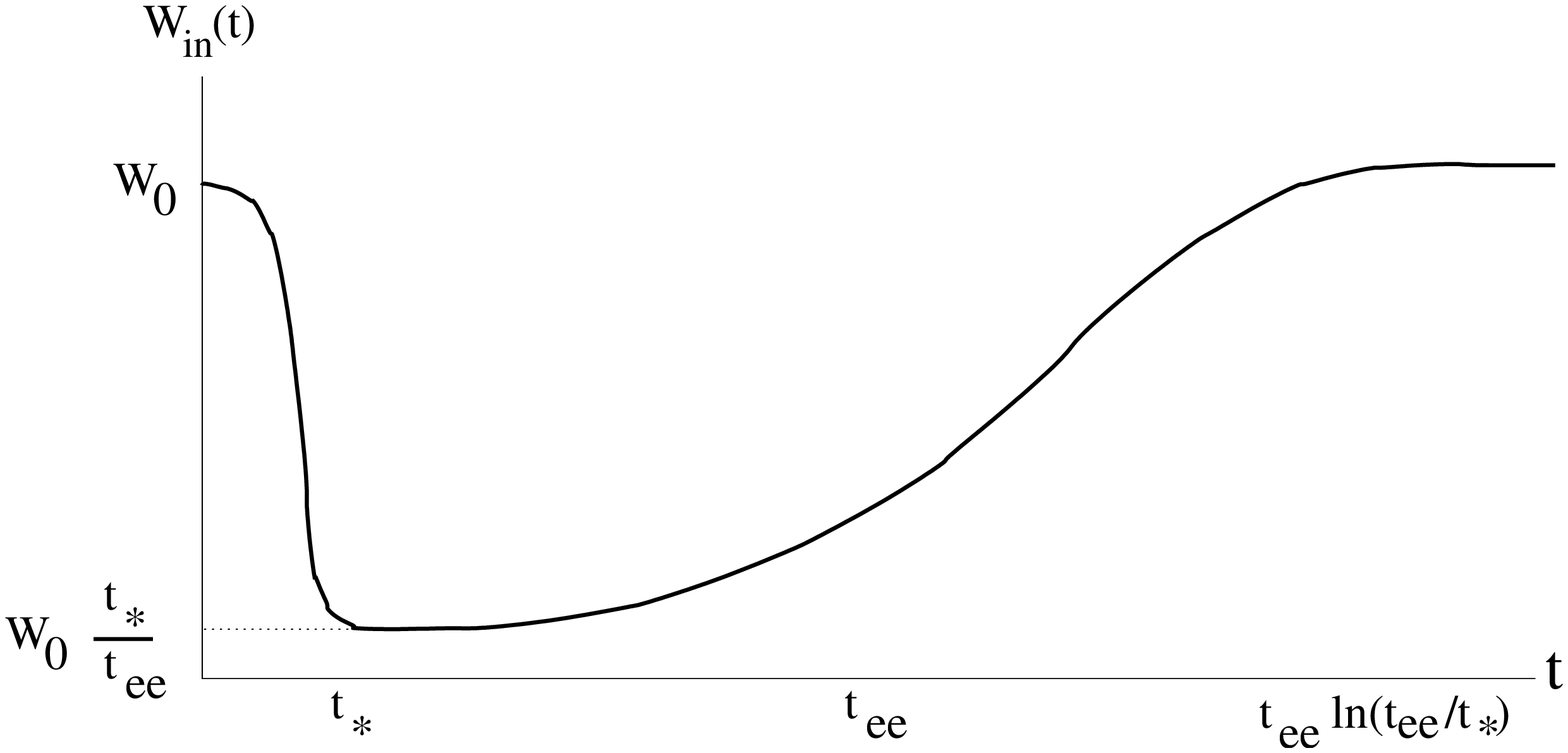}
\caption{Absorption rate vs time in a closed system with 
electron-electron interaction.}
\end{figure}
\noindent
The main conclusion of the qualitative picture of Ref.\cite{Bas}
described above is that the dynamic localization is a {\it 
transient phenomenon} that is destroyed at long enough times by an 
arbitrary small electron interaction.

Note, however, that the consideration of Ref.\cite{Bas} is based on 
the Fermi Golden Rule. The applicability of this description is 
known \cite{AGKL} to be restricted in quantum dots by the region of 
relatively large
electron energies $T_{\rm eff}\gg \sqrt{E_{Th}\delta}$.
At smaller energies the localization in the Fock space should take 
place with the effect of vanishing $1/t_{\rm ee}$. Therefore it 
is 
likely that
at $T_{*}\sim \omega \Gamma/\delta < \sqrt{E_{Th}\delta}$ the 
dynamic localization survives electron-electron interaction.
\section{Dynamic localization in an open quantum dot and the shape 
of the Coulomb blockade peak.}
The transient character of DL and difficulty to 
realize a closed system in condensed matter physics because of the 
phonon heat transport leaves only one option for possible 
observation of DL: the steady state in an {\it open} system under 
periodic pumping. In an open system Eq.(\ref{AvAR}) should be 
generalized to include the dephasing due to electron escape into 
leads along with dephasing by inelastic electron collisions 
considered in the previous section:
\begin{equation}
\label{Win-gen}
W_{\rm 
in}=W_{0}\,\frac{\gamma_{\varphi}t_{*}}{1+\gamma_{\varphi}t_{*}},
\end{equation}
where the energy-dependent dephasing rate is given by:
\begin{equation}
\label{gamma-phi}
\gamma_{\varphi}=\gamma_{\rm esc}+\frac{1}{t_{\rm ee}(T_{\rm 
eff})}.
\end{equation}
From Eq.(\ref{Win-gen}) it follows that in order to observe any 
deviation from the classical absorption rate $W_{0}$ the escape 
rate $\gamma_{\rm esc}$ must be much smaller than the inverse 
localization time $1/t_{*}\sim \delta^{2}/\Gamma$. Then we obtain
the necessary condition for DL:
\begin{equation}
\label{nec}
\frac{\gamma_{\rm esc}}{\delta}\ll \frac{\delta}{\Gamma}\ll 1.
\end{equation}
The last inequality Eq.(\ref{ga-de}) is the basic approximation 
adopted 
in this
study of DL.

Eq.(\ref{nec}) shows that the DL effects considered could be 
observed only provided that the quantum dot is {\it almost closed},
i.e. $\frac{\gamma_{\rm esc}}{\delta}\ll 1$. This is exactly the 
condition where the Coulomb blockade effects \cite{Alei-rep} are 
relevant. Under this condition the low-temperature linear dc 
conductance 
of the dot is 
strongly peaked near the value of the gate voltage $V_{g}=V_{c}$ 
where the states
with $N$ and $N+1$ electrons on the dot are nearly degenerate. 
At equilibrium conditions where the electron energy distribution
inside the dot is Fermi-Dirac with the same temperature as in the 
leads, the width of the peak is proportional to the temperature
and thus can be used as a thermometer. In the case where the dot is 
subject to time-dependent perturbation, the electron energy 
distribution inside the dot can be drastically different from that
in the leads and a non-trivial question arises which of these two 
distributions affect the linear dc conductance. The answer for an
open dot where the Coulomb blockade effects play no role is given 
in Refs.\cite{Kanzieper, Vavilov}. It appears that in this case the 
linear dc conductance is sensitive {\it only} to the electron 
energy 
distribution {\it in the leads} and thus cannot be used to probe 
energy dynamics in the dot. However, with the Coulomb interaction 
inside the 
dot taken into account and for the case $T_{\rm eff}\gg \delta$  
the linear dc conductance becomes sensitive to the effective 
electron temperature $T_{\rm eff}$ inside the dot. Below we 
consider only the extreme case of the Coulomb blockade.
We also assume that the inelastic electron interaction is
sufficiently strong to produce the Fermi-Dirac form of the electron 
energy distribution inside the dot, though with the temperature
$T_{\rm eff}$ much larger than that in the leads. We concentrate on 
the qualitative changes in the Coulomb blockade peak shape 
characteristic of the dynamic localization in the approximation of 
sequential tunneling \cite{Alei-rep}. More detailed consideration
including the role of inelastic co-tunneling and cooling by 
phonons is 
presented in Ref.\cite{KrBas}.

The steady state of a dot under ac perturbation is determined by 
the global energy balance $W_{\rm in}=W_{\rm out}$, where the 
pumping of 
energy by the ac perturbation is governed by Eq.(\ref{Win-gen}) 
and $W_{\rm out}$ is the cooling rate. The most interesting case
corresponds to a situation where the main cooling mechanism is
due to electron escape into cold leads. Denoting by $U=V_{g}-V_{c}$
the detuning of the gate voltage from the peak center and 
introducing the parameter $x=U/(2T_{\rm eff})$ we obtain 
\cite{KrBas}
for the case of leads maintained at a temperature $T_{0}\ll T_{\rm 
eff}$:
\begin{equation}
\label{cool}
\frac{W_{\rm out}(U)}{(\gamma/\delta)\, T_{\rm 
eff}^{2}}=\frac{\pi^{2}}{12}-x^{2}+
\frac{2x}{\ln(2\cosh x)}\,\int_{0}^{x}\xi\,\tanh \xi\,d\xi,
\end{equation}
where $\gamma=2\gamma_{\rm esc}(U=0)$. 
\begin{figure}[ht]
\vskip.2in
\includegraphics[width=.7\textwidth]{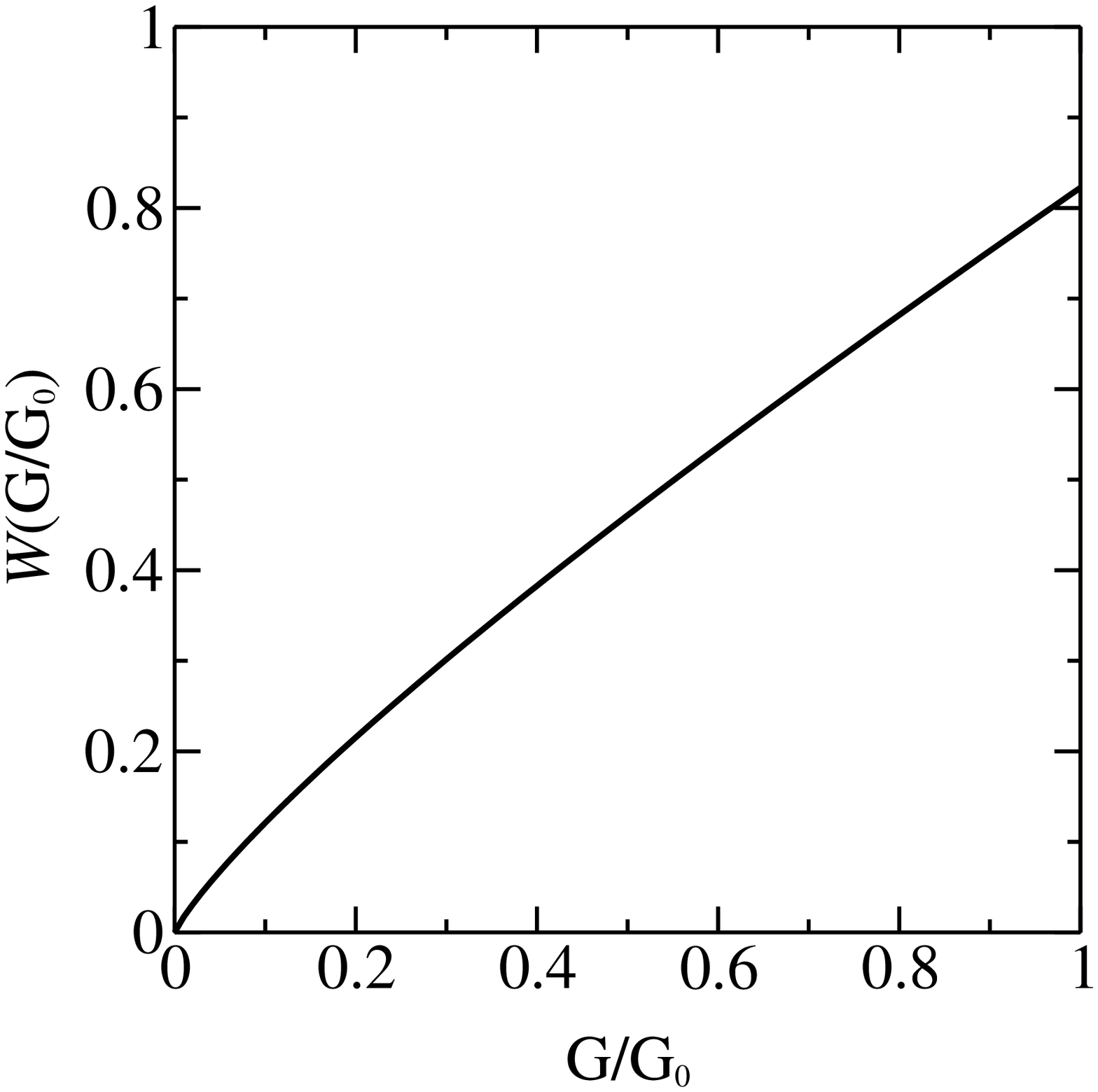}
\caption{The function ${\cal}(G/G_{0})$ in Eq.(\ref{Wied-Fr}).}
\end{figure}
In the same approximation we obtain for the linear dc conductance 
of the dot:
\begin{equation}
\label{dc-cond}
\frac{G(U)}{G_{0}}=1-\frac{x\,\tanh x}{\ln(2\,\cosh x)}.
\end{equation}
Remarkably, the r.h.s. of Eqs.(\ref{cool},\ref{dc-cond})
depends only on the single variable $x$, so that one can
establish a relationship similar to the Wiedemann-Franz law:
\begin{equation}
\label{Wied-Fr}
W_{\rm out}=\left (\frac{\gamma}{\delta}\right)\,T^{2}_{\rm 
eff}\,{\cal W}(G/G_{0}).
\end{equation}
Now, let us assume that
\begin{equation}
\label{Cond1}
\gamma_{\varphi}t_{*}\ll 1,
\end{equation}
and
\begin{equation}
\label{Cond2}
\frac{\gamma_{\rm esc}(U)}{\delta}\ll \frac{T^{2}_{\rm 
eff}}{E_{Th}^{2}},
\end{equation}
where the escape rate is given by
\begin{equation}
\label{EscRate}
\frac{\gamma_{\rm 
esc}}{\gamma}=\frac{1}{2}-\frac{|x|}{2\ln(2\cosh x)}.
\end{equation}
Then we find from Eqs.(\ref{SIA},\ref{Win-gen}):
\begin{equation}
\label{IN}
W_{\rm in}=T_{\rm eff}^{2}\,\left(\frac{T_{*}}{E_{Th}}\right)^{2}.
\end{equation}
\begin{figure}[ht]
\vskip.2in
\includegraphics[width=.8\textwidth]{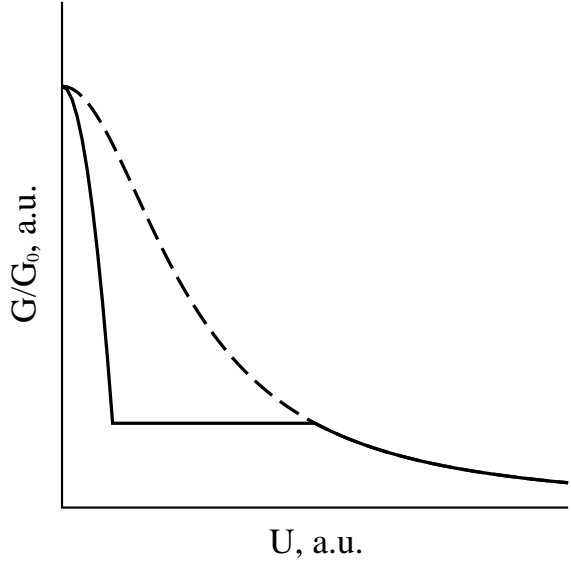}
\caption{A sketch of the Coulomb blockade peak shape in the
dynamic
localization regime without taking into account the phonon
cooling (solid line): at small~$U<U_{\rm{min}}$ the dephasing
is dominated by the electron escape (peak), at larger~$U$ --
by electron-electron collisions (plateau), and finally, at
$U>U_{\rm{max}}$ the cooling is insufficient, the dynamic
localization is destroyed, and the dot is in the Ohmic regime.
The Ohmic curve is also shown for reference by the dashed line.}
\end{figure}
One can see that due to the fact that both
the cooling rate $W_{\rm out}$ given by
Eq.(\ref{Wied-Fr}) and the energy pumping  rate $W_{\rm in}$ given
by Eq.(\ref{IN}) are proportional to $T^{2}_{\rm eff}$, the
effective electron temperature drops out of the global
energy balance $W_{\rm in}=W_{\rm out}$. As the result, the dc
conductance of the dot is remarkably independent of the gate
voltage:
\begin{equation}
\label{plato}
\frac{G}{G_{0}}\sim
\left(\frac{T_{*}}{E_{Th}}\right)^{2}\,\frac{\delta}{\gamma}.
\end{equation}
Note that since $G/G_{0}<1$ such a {\it plateau} solution exists
only when
\begin{equation}
\label{COND}
\frac{\gamma}{\delta}> \left(\frac{T_{*}}{E_{Th}} \right)^{2}.
\end{equation}
At the same time for large enough detuning $U>U_{\rm min}\sim
T_{*}$ the condition Eq.(\ref{Cond2}) is fulfilled because the
escape rate $\gamma_{\rm esc}(U)$ decreases
and the effective temperature increases with increasing $U$.
However, with $U$ and $T_{\rm eff}$ further increasing the
DL regime Eq.(\ref{Cond1}) is destroyed and $G(U)/G_{0}$ starts to
decrease. This happens at $U>U_{max}\sim
E_{Th}\,\sqrt{\frac{\delta}{\Gamma}}$.
The sketch of the the $G(U)$ dependence is shown in Fig.7.
\section{Summary}
In conclusion, we have considered the manifestation of dynamic 
localization in a solid-state system -- a quantum dot under 
time-dependent perturbation. We have shown that in a closed dot
without electron interaction the dynamic localization is possible
and we developed the theory of weak dynamic localization for an 
arbitrary time-dependent perturbation. Then we considered a 
realistic case of a dot with electron-electron interaction weakly 
connected to leads. We have demonstrated that if 
the dot is subject to a periodic 
perturbation the dynamic localization can result in a plateau at
the tail of the Coulomb blockade peak in the linear dc conductance 
vs the gate voltage and we established the 
conditions under which such a plateau arises.

\begin{acknowledgments}
The author wish to thank D.M.Basko and M.A.Skvortsov for a 
collaboration on the problems discussed in the article
and B.L.Altshuler, I.V.Lerner and V.I.Falko for illuminated 
discussions.
\end{acknowledgments}

\end{document}